**Title**

**Generalized method of images and reflective color generation from ultra-thin multipole resonators**

**Authors**

S. Q. Li,[1]*† W. Song,[2,3] M. Ye,[1] K. B. Crozier[1,2]*

**Affiliations**

1. Department of Electrical and Electronic Engineering, University of Melbourne, VIC 3010 Australia

2. School of Physics, University of Melbourne, VIC 3010 Australia

3. Department of Materials Science and Engineering, Huazhong University of Science and Technology, Wuhan 430030, PR China

*S. Q. Li: lishiqia@gmail.com; K. B. Crozier: Kenneth.crozier@unimelb.edu.au

† Present Address: Data Storage Institute, 2 Fusionopolis Way, #08-01 Innovis Tower, Singapore 138634

**Abstract**

The multipole expansion has found limited applicability for optical dielectric resonators in inhomogeneous environment, such as on the surface of substrates. Here, we generalize the method of images to multipole analysis for light scattering by dielectric nanoparticles on conductive substrates. We present examples illustrating the physical insight provided by our method, including selection rules governing the excitation of the multipoles. We propose and experimentally demonstrate a new mechanism to generate high resolution surface color. The dielectric resonators employed are very thin (less than 50 nm), i.e. similar in thickness to the plasmonic resonators that are currently being investigated for structural color. The generalized method of images opens up new prospects for design and analysis of metasurfaces and optical dielectric resonators.

**MAIN TEXT**

**Introduction**

Optical devices comprising microstructures arranged on a surface to perform an optical function normally achieved with a bulk optical element have a long history, with an early example being the Fresnel zone plate.[1] In recent years, the constituent building blocks of these devices have gone from the micro-scale down to the nano-scale. Concepts from the field of optical metamaterials[2,3] have been adopted and extended, and these devices are now often termed metasurfaces. A key development was the demonstration that planar arrangements of resonant metallic nanostructures could be used to imprint a prescribed phase profile upon a transmitted beam, thereby leading to interesting functionalities such as anomalous refraction[4,5] and vortex beam generation[4]. While many different optical functions have been demonstrated with metallic nanostructures,[6-8] attention has recently turned to dielectrics as the building blocks for metasurfaces, due to significantly lower loss. Dielectric metasurfaces can in general be classified into non-resonant and resonant designs.

An example of the non-resonant approach is the work of Schonbrun et al[9], who necessary demonstrated metasurface lenses comprising arrays of elliptical silicon to produce different phase accumulations for light polarized along different axes. Other non-resonant metasurfaces include circular polarization beam-splitters,[10] holograms and vector beam generators,[11] linear polarization beam-splitter pixels,[10] high numerical aperture lenses,[12] and retro-reflectors.[13] A challenge of this approach is that the nanopillars need to be relatively tall (on the order of the wavelength divided by the refractive index). By contrast, in the resonant approach, the dielectric building blocks are generally much thinner.[14] In this method, it is the electric dipole (ED) and magnetic dipole (MD, and possibly higher order) resonances[15, 16] of dielectric nanoparticles such as nanodisks that control the phase.[17, 18] By appropriate design, the ED and MD resonances can occur at the same frequency, allowing the metasurface in principle to have a phase shift of $2\pi$ and near-unity transmission.[19] The latter results from the ED and MD interfering constructively and destructively in the forward and backward directions, respectively.[20, 21] A number of interesting applications of metasurfaces based on resonant dielectric nanoparticles have been demonstrated. Examples include surface-enhanced spectroscopy,[22, 23] high resolution colour printing,[24, 25] magnetic mirrors,[26] non-linear optics,[27] and the ability to generate the Brewster effect with any polarisation and at any angle of incidence.[28]

Despite the successes of metasurfaces based on resonant dielectric nanoparticles,[29] an important challenge remains. In the example applications listed above, the dielectric nanoparticles were fabricated on dielectric substrates (or on thick films) with lower refractive indices. In some cases, the nanoparticles were coated with an overlayer whose index matched that of the substrate, with the goal of embedding the nanoparticles in a homogeneous environment.[30] In these situations, the multipole expansion can be readily applied and is very helpful in elucidating the underlying optical physics. Physical interpretation can be much more difficult for highly inhomogeneous environments, e.g. when the resonant dielectric nanoparticle is on a metal substrate. The latter is of particular importance for two reasons. First, there are numerous established and potential applications for metal mirrors and for other types of reflective optics[13, 31-34] and the inclusion of resonant dielectric nanoparticles could present interesting new opportunities. Second, the ability of dielectric metasurfaces to control the propagation and emission of light could be useful in many optoelectronic devices. Such devices however are generally quite inhomogeneous, comprising multiple layers including metals. For these and other reasons, experimental and theoretical studies have recently appeared concerning resonant dielectric above metal films.[31, 35-37] Here, we derive a generalized formulation for multipole analysis of dielectric resonators above a substrate, based on the classical method of images.[38, 39] The simplicity of this formulation facilitates physical interpretation and the design of dielectric resonators. As an example, we employ it to demonstrate a new method for structural colour patterning at high resolution using silicon resonators with less than 50 nm thickness.

**Theoretical Development**

The method of images is an established technique for the analysis of electromagnetic field due to electrostatic charges and dipole radiations near conductive surfaces.[38, 39] For example, it simplifies the calculation of radiation from an infinitesimal dipole on a perfect electric conductor (PEC) to the problem of finding a virtual source (image dipole) that accounts for the reflection. The combination of the actual and virtual sources form an equivalent system that gives the same radiated field above the PEC as the actual system. The image dipole generates fields that may interfere constructively or destructively with those generated by the actual source dipole as the direction of the image dipole depends on both the nature of actual source dipole (i.e. whether it is magnetic or electric) and its orientation (e.g. whether it is points vertically or horizontally). The scattering of light from sub-wavelength objects can be approximated as dipole radiation. For nanoparticles with high refractive index and that are too large to be treated in the electrostatic limit, the scattering cross-section may have contributions from modes with orders higher than dipole, such as quadrupoles and octupoles. The method of images thus should be generalized to describe multipoles. To achieve this, we define a new set of multipole coefficients to describe the scattering behaviour of objects on conductive surfaces. We term these the **effective multipole coefficients**:

$$a_{eff}^{e}(l,m) = a^{e}(l,m) - a^{e}(l,-m) \qquad (1)$$

$$a_{eff}^{m}(l,m) = a^{m}(l,m) + a^{m}(l,-m) \qquad (2)$$

, where $a^{e}(l,m)$ and $a^{m}(l,m)$ are electric and magnetic multipole coefficients calculated by integrating the polarisation current (= $\varepsilon_0 \chi_e \mathbf{E}$) in the scattering objects, and $l$ and $m$ are the order and the degree of the multipole under consideration.[40] This approach imposes no restrictions on the shape and orientation of the scatterer. It also imposes no restrictions on the type and direction of the incident wave. A detailed derivation along with some examples is presented in the Supplementary Information (SI). The partial scattering cross-section of an order $l$ is calculated in the usual way as,

$$C^{e,m}(l) = \frac{\pi}{2k^2} \sum_{m=-l,}^{l} (2l+1) |a_{eff}^{e,m}(l,m)|^2 \qquad (3)$$

, where $k$ (= $\omega/c$) is the wavenumber of the interacting time-harmonic wave. This equation differs from the partial scattering cross-section calculated with the traditional multipole coefficients only by the factor of 2 in the denominator of the pre-factor, as it is associated with scattering over a hemisphere.

In Figure 1, we show the spectra of the partial scattering cross-sections for the first three orders of $l$, corresponding to dipoles, quadrupoles, and octupoles respectively. Two cases are shown. The first is that of plane wave illumination of a silicon sphere (Figure 1(a)). In the second, a plane wave is normally-incident on a silicon hemisphere on a PEC (Figure 1(b)). The hemisphere of Figure 1(b) has the same radius as the sphere of Figure 1(a). In each case, the total scattering cross-section is also shown, calculated by integrating the Poynting vector associated with the scattered field over a surface enclosing the sphere or hemisphere. We find that for both cases these total scattering cross-section spectra (blue curves of Figure 1(a) and (b)) are in complete agreement with spectra consisting of the summation of the first three orders (coloured curves of Figures 1(a) and (b)).

Comparison between the partial scattering cross-sections from the isolated sphere (Figure 1(a)) and those from the hemisphere on the PEC (Figure 1(b)) leads to some interesting observations. First, the magnetic dipole (MD) spectra of the sphere and hemisphere-on-PEC have the same peak positions. It is notable that the magnetic dipole resonant peak position is unaffected by replacing half of the sphere with a PEC. Second, the electric dipole (ED) response vanishes for the hemisphere-on-PEC configuration while the magnetic dipole does not. This is consistent with the traditional method of images, in which a horizontal source electric dipole is associated with an image electric dipole that is antiparallel with it.[38] Similarly, a horizontal source magnetic dipole is associated with an image magnetic dipole that is parallel with it. Third, the second order resonances, i.e. the electric and magnetic quadrupoles, in the isolated sphere case are completely modified for the hemisphere on PEC case. In particular, the electric quadrupole (EQ) of the hemisphere-on-PEC case has the response of the magnetic octupole (MO) of the isolated sphere. The response of the electric octupole (EO) of the isolated sphere is incorporated in the response of the MD of the hemisphere-on-PEC, broadening the scattering peak that is in the spectral range 800-900 nm. Similarly, the resonance in the spectrum of the magnetic quadrupole (MQ) of the isolated sphere is completely absent in the MQ of the hemisphere-on-PEC case, whose spectrum is rather flat.

This selection and re-ordering of multipoles can be understood from the symmetry of radiation field and the boundary condition on the substrate. For example, the peaks in the spectra of the three modes ED, EQ, and MQ of the free sphere are no longer present in any of the spectra of the hemisphere-on-PEC case because their field distributions are incompatible with the PEC boundary condition. A closer examination of the EQ of the hemisphere-on-PEC however reveals has a slightly asymmetric line-shape, which we interpret as being due to it retaining some of the response of the EQ of the free sphere. We interpret the incorporation of the EO mode (of the free sphere) into the MD mode (of the hemisphere-on-PEC) as arising from the fact that the original EO mode (of the free sphere) is partially compatible with the boundary condition. This leads to a cancellation of the incompatible components, with the remaining components being incorporated in the MD mode. Further discussions on the conversion and eliminations of octupoles and quadrupoles are provided in the SI (see Figure S-6 and the discussion therein).

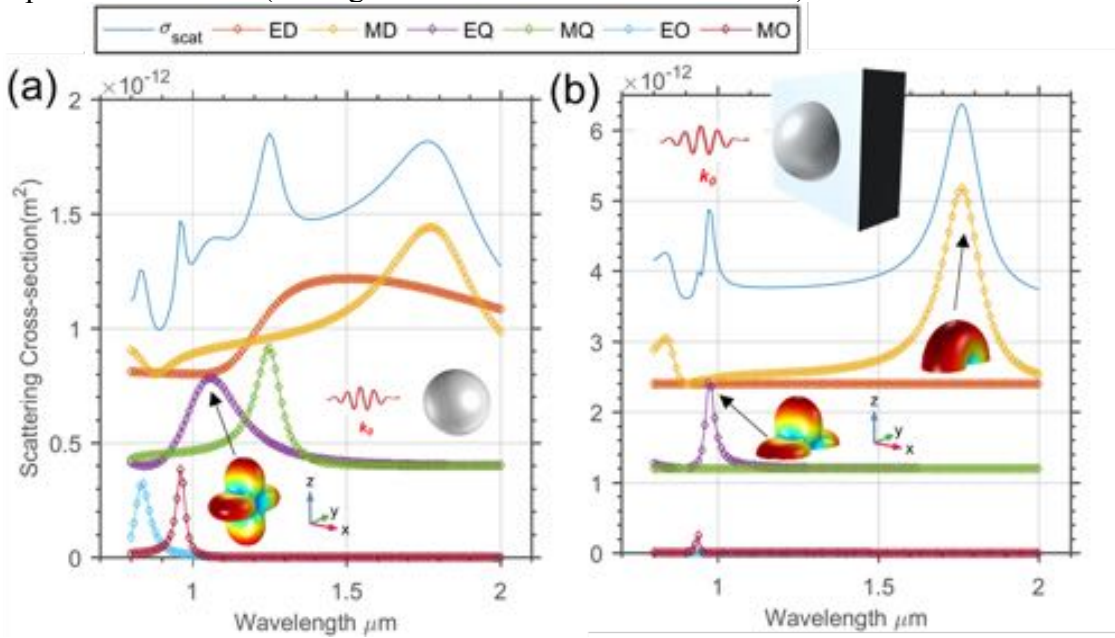

*Figure 1. Generalized method of images for multipole scattering by dielectric nanoparticles on conductive substrates.*

*(a) Partial scattering cross-section spectra of multipolar components for silicon sphere (230 nm radius) with plane wave illumination, as found by integration of polarisation current. Total scattering cross section (blue curve), as found by integration of Poynting vector associated with scattered fields. Curves have been shifted vertically for clarity of presentation (by $4(l-1)\times10^{-13}$ $m^2$). Electric dipole, magnetic dipole, electric quadrupole, magnetic quadrupole, electric octupole and magnetic octuopole are denoted ED, MD, EQ, MQ, EO, and MO, respectively. (b) Same as for panel a, except for hemisphere-on-PEC and with partial scattering cross-sections found by generalized method of images (Equations 1-3). Hemisphere has radius of 230 nm. Curves have been shifted vertically for clarity of presentation (by $12(l-1)\times10^{-13}$ $m^2$). Insets show schematic illustration of illumination of*

*sphere (panel a) and hemisphere (panel b). Insets also show simulated radiation patterns of EQ (panel a), and of EQ and MD (panel b).*

The multipole mode selection by PEC transforms a rather rich and broad scattering spectrum (solid blue curve of Figure 1(a)) to one with well-separated peaks (solid blue curve of Figure 1(b)). Indeed, there is a broad wavelength range of low scattering for the hemisphere-on-PEC case. This wavelength range with low scattering for the single hemisphere-on-PEC would result in a wavelength range of high reflectance for an ensemble of hemispheres-on-PEC. This presents a new mechanism for controlling the reflection spectrum of a surface that could be applied in colour printing. The colour that the surface appears could be readily controlled by appropriate choice of nanoparticle size. This leads to the question of how this method for colour printing should be realised in practice. An important consideration is that of manufacturing feasibility. It would be difficult to fabricate an array of silicon hemispheres. Using standard processes however, silicon nanoscale disks and cylinders can be fabricated with excellent control of critical dimensions. It is furthermore known that the resonances supported by such shapes are very similar to those of spheres.[17] Another consideration is what material to use for the substrate, whose response should approximate that of a PEC. We choose aluminium because of its high plasma frequency (i.e. small skin depth across the entire visible region) and its low cost. In the remainder of this paper, we investigate the optical properties of silicon nano-cylinders and nano-disks on conductive substrates. We consider the scattering of light by silicon nano-cylinders on a PEC using our generalised method of images (Eq. 1 to 3). We then show how this method should be modified when the substrate is not a perfect conductor, but rather a realistic metal. We then analyse the optical properties of silicon nano-disks on an aluminium substrate, and experimentally demonstrate their application to colour printing. Note that to be consistent with the literature, in this paper we refer to silicon as a dielectric material to distinguish it from a metal.

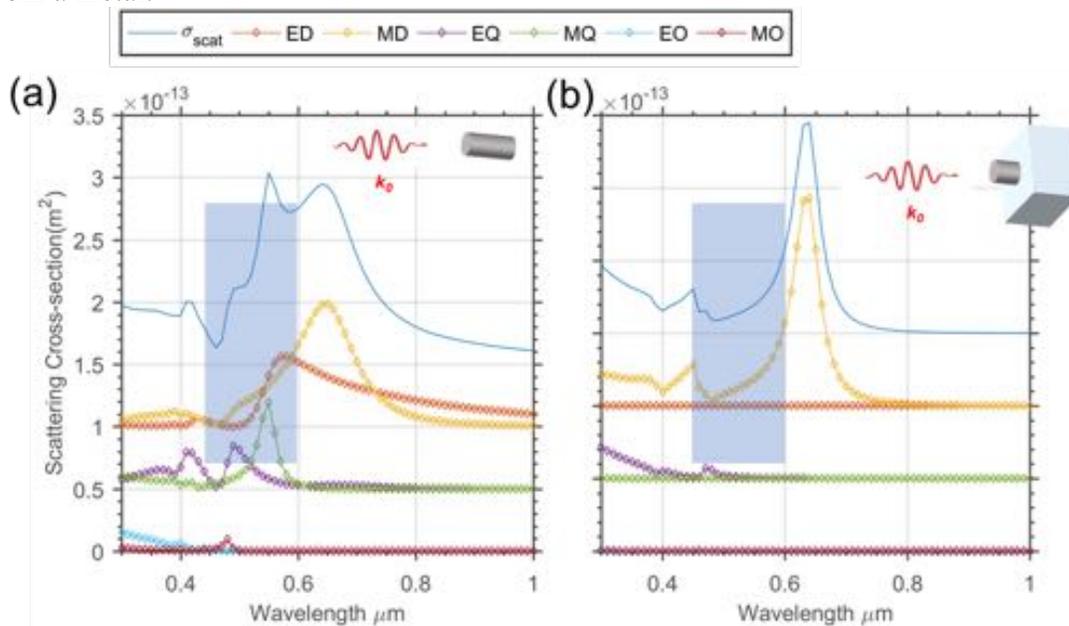

*Figure 2. Generalized method of images analysis of light scattering by a dielectric nano-cylinder in free space and on a PEC. (a). Partial scattering cross-section spectra of multipolar components for silicon nano-cylinder (height: 100 nm, radius: 75 nm) with plane wave illumination. Total scattering cross section spectrum shown as blue curve. Curves have been shifted vertically for clarity of presentation (by $5(l-1) \times 10^{-14}$ $m^2$). (b). Partial scattering cross-section spectra of multipolar components for silicon nano-cylinder (height: 50 nm, radius: 75 nm) on PEC, with plane wave illumination. Total scattering cross section*

*spectrum shown as blue curve. Curves have been shifted vertically for clarity of presentation (by (l-1)×10⁻¹³ m²).*

We now apply our generalized method of images to the analysis of a silicon nano-cylinder in free space (Figure 2(a)). It can be seen that the first three orders of multipole resonances are sufficient to account for total scattering cross section in the visible region (400 nm to 700 nm). The strengths of the resonances decrease with order. We next analyse light scattering by a silicon nano-cylinder on a PEC (Figure 2(b)), whose height is half that of the nano-cylinder in free space of Figure 2(a). It can be seen that the suppression of the quadrupole resonances and electric dipole resonances creates a spectral region with low scattering, which is indicated with blue shading in Figure 2. This region is ~150 nm wide, and offers interesting possibilities for reflected colour.

To apply our generalised method of images at visible wavelengths, we must take into account the fact that metals are not PECs in this spectral range. Real metals have finite and complex refractive indices, with the latter resulting in absorption of the fields created by the multipoles. To model this appropriately, we introduce a parameter called image strength, denoted by $\Gamma$, to modify Eqn. (1) and (2). These equations thus become

$$a^e_{eff}(l,m) = a^e(l,m) - \Gamma a^e(l,-m) \qquad (4)$$

$$a^m_{eff}(l,m) = a^m(l,m) + \Gamma a^m(l,-m) \qquad (5)$$

, where $\Gamma = \dfrac{\varepsilon_s - \varepsilon_1}{\varepsilon_s + \varepsilon_1}$, and $\varepsilon_s$ and $\varepsilon_1$ are the permittivities of the substrate and of the surrounding region (e.g. air).

**Experimental**

We now describe our experimental realization of colour printing with dielectric resonators on conductive substrates. We begin by considering the optical properties of silicon nano-disks on aluminium substrates by simulations and experiments. In these simulations, the complex refractive index for aluminium is taken from the compilation of Palik[41], while that for silicon is based on the values we measure experimentally by ellipsometry (Figure 3(a)). This plot also includes the complex refractive index of amorphous silicon from the literature.[42] We observe that the real ($n$) and imaginary ($k$) parts of the refractive index of the silicon we deposit by electron-beam evaporation (EBE) are much smaller than those of the reference value for amorphous silicon. We attribute the reduced indices to be the result of our film having a low packing density. The importance of sample preparation upon the optical properties of amorphous silicon was recognized by Pierce and Spicer.[42] In Figure 3(b), we plot the partial scattering cross sections of the EQ and MD modes of the silicon nano-disk on aluminium, calculated using our generalized method of images (Equations 4-5). The total scattering cross section is also shown. The MD resonance of Figure 3(b) is broader than it would be, had we employed the optical properties of a-Si as reported by Pierce and Spicer rather than those of our EBE-deposited material. It can be seen nonetheless that the total scattering cross section (Figure 3(b)) displays a pronounced trough between the resonant peaks of the MD and EQ. This is created by the conductive substrate eliminating the ED, and modifying the nature of the EQ and MQ modes. The radiation patterns at the wavelengths of $\lambda = 380\ nm$ and $660\ nm$ are plotted in the inset. The pattern at $\lambda = 380\ nm$ is consistent with the upper half of the four-lobed pattern that one would expect for electric quadrupoles ($l = 2,\ m = \pm 1$). The bottom half of the pattern one would expect for an electric quadrupole is of course absent due to the presence of the conductive substrate. Similarly, the pattern at $\lambda = 660\ nm$ has a half-donut shape, which is consistent with it being the upper half of the donut pattern that one would expect for magnetic dipoles ($l =1$, $m = \pm 1$), shown as the inset in Figure 1(b). It can also be seen that the radiation intensity along the surface is small compared to that occurring in Figure 1(b). This is due to resistive losses in the aluminium that occur when propagation is along the surface.

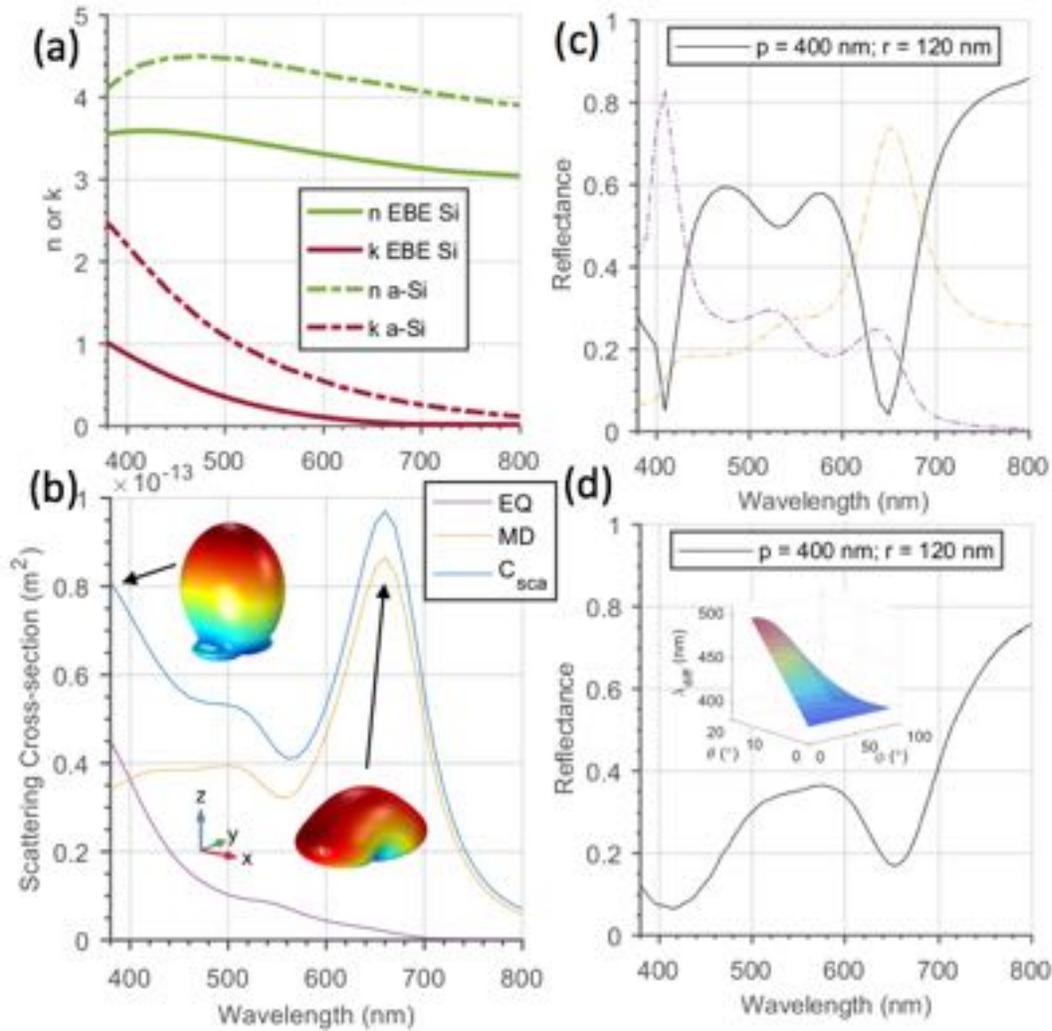

*Figure 3. **Optical properties of silicon nano-disks on aluminium substrates: simulations and experiments.** (a). Complex refractive indices vs wavelength of amorphous silicon, as measured experimentally for films we deposit by electron beam evaporation (EBE Si) and as previously reported in the literature (a-Si).[42] (b) Partial scattering cross-section spectra of electric quadrupole and magnetic dipole terms for isolated silicon nano-disk on aluminium. Total scattering cross section ($C_{sca}$) is also shown. Radiation patterns that occur at peaks in scattering cross section spectra are also plotted, with the orientation of the coordinate axes (xyz) as shown. Illumination is from a plane wave propagating along the z axis (-z direction), whose electric field is along the y axis. (c). Simulated reflectance spectrum (black curve) of square array (period 400 nm) of silicon nano-disks (radius 120 nm and height 45 nm) on an aluminium substrate. Absorption within silicon and within aluminium vs wavelength are plotted as purple and yellow curves, respectively. (d) Experimentally measured reflectance spectrum of square array (period 400 nm) of silicon nano-disks (radius 120 nm) on aluminium. Inset: calculated wavelength at which first diffraction order transitions from evanescent to propagating, as a function of azimuthal $\varphi$ and elevation $\theta$ angles of incident plane wave.*

We now consider the case in which the silicon nano-disks are formed in an array. The reflection spectrum simulated for a square array (period 400 nm) of silicon nano-disks (radius 120 nm, height 45 nm) on an aluminium substrate is shown as Figure 3(c). It can be seen that there are dips in the reflectance at wavelengths of $\lambda \approx 410\ nm$ and $650\ nm$. It can also be seen that these dips are associated with peaks in the absorption spectra of the silicon nano-disks (purple curve) and of the aluminium (yellow curve), which are also plotted in Figure 3(c). As we describe in further detail below, we fabricate an array of silicon disks on an aluminium substrate. We measure the reflection spectrum of this sample (Figure 3(d)). It can be seen that the experimentally-measured spectrum (Figure 3(d)) exhibits dips whose positions are very close to the predictions of simulations (Figure 3(c)). There are some differences between the spectra however. In experiments, the longer-wavelength dip has a minimum reflectance value of ~0.17, while simulations predict ~0.04. For the spectral region between the dip features, the measured reflectance is smaller than the predictions of simulations. These differences may be in part due to the fact that the simulations are performed for normal incidence illumination, while in the experiments a microscope objective with a numerical aperture (NA) of 0.1 is used. This NA corresponds to an illumination angle, i.e. elevation angle $\theta$, of up to 15°. The reflection spectrum is expected to be dependent on the angle of illumination for two reasons. First, the optical response of an isolated silicon nano-disk (on aluminium) will in general vary with angle of incidence. Second, for a silicon nano-disk array, interactions between nano-disks play an important role in the optical response. These interactions are dependent on the angle of incidence. As an example, consider the wavelength at which light near normal incidence on the array is coupled into a diffraction order that runs along the plane of the substrate. This leads to a pronounced interaction between nanoparticles and has also been termed the "lowest order lattice resonance". As shown in the inset of Figure 3(d), this wavelength varies with the elevation and azimuthal angles of the incident light. It matches the array period (400 nm) at normal incidence, and is larger at other angles. In addition to the angle of illumination, the differences between simulation and experiment can be attributed to surface roughness of the aluminium surface (~6 nm, as determined by atomic force microscopy).

Further insight into the optical response of the silicon nano-disk array can be obtained by considering the field distributions that occur at the wavelengths of the dips of the reflection spectrum (Figure 3(c)). As discussed above, at these wavelengths ($\lambda = 410\ nm$ and $650\ nm$), the absorption spectra of the silicon nano-disk and of the aluminium film exhibit peaks. We plot the field distributions at these wavelengths in Figure 4. It can be seen that for illumination at $\lambda = 410\ nm$, the field peaks at the corners of the disks (Figure 4(a) and 4(b)). Though it is a little difficult to see from Figure 4(a) and 4(b) due to the very strong fields at the corners, the field is also appreciable inside the disks, which results in absorption. These field patterns are consistent with quadrupole resonances. These resonances are enhanced due to the lattice resonance occurring at around the same spectral position ($\lambda = 400\ nm$). For illumination at $\lambda = 650\ nm$, it can be seen that the field at the boundary between the nano-disk and the aluminium substrate is appreciable (Figure 4(c) and 4(d)). This is consistent with the resonance producing a peak in the absorption spectrum of the aluminium substrate (Figure 3(c)).

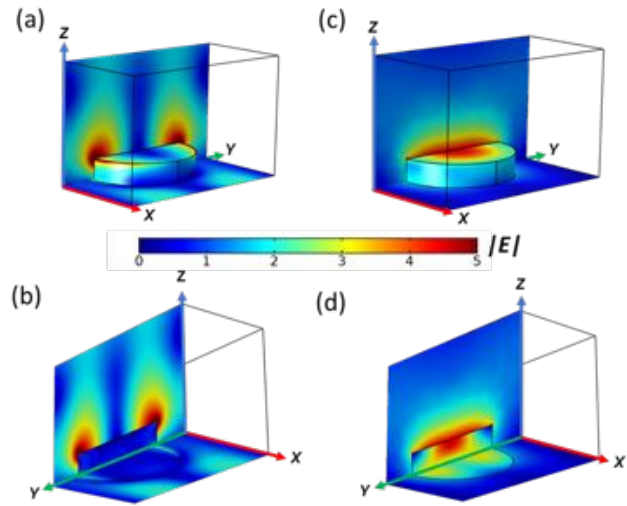

*Figure 4. Simulated electric field strength distribution when silicon nano-disk array (on Al) is illuminated at a wavelength of 410 nm. Front and back views are shown as panel (a) and (b). (c) and (d) are the distributions for illumination at a wavelength of 650 nm.*

We now describe the fabrication method for the silicon nanodisk array on aluminium. Key steps are shown in Figure 5(a). Fabrication starts with a silicon wafer (4 inch, (100) orientation) onto which aluminium (100 nm) and then amorphous silicon (50 nm) films are deposited by e-beam evaporation (EBE). The wafer is then diced with a wafer saw into square samples (15 mm side length). E-beam resist (ZEP520A, Zeon Chemicals) is then spun onto the sample which is then baked, exposed using standard e-beam lithography (EBPG5000plusES, Vistec), and developed. These steps result in an array of disks in the resist, which is transferred to the underlying silicon by inductively-coupled reactive ion etching using $SF_6$ and $C_4F_8$ gases (Plasmalab100 ICP380, Oxford Instruments). The resist is then removed by soaking the sample in ZDMAC remover (Zeon Chemicals) for 5 minutes followed by rinsing with iso-propanol. A scanning electron microscope (SEM) image of one of the finished arrays is shown as Figure 5(b). The disks in this array have radii of 90 nm (design value of e-beam lithography step) and are in square arrays of period 300 nm. Atomic force microscopy (Figure 5(c)) confirms that the disks have heights of ~ 50 nm.

A bright-field optical microscope image of ten disk arrays is shown as Figure 5(d). This image is obtained with illumination from a tungsten lamp. The microscope objective lens has an NA of 0.15. Each array consists of a square lattice of silicon nano-disks, whose radii range from 75 to 150 nm. The larger arrays have overall extents of 200 by 200 $\mu m$, and the smaller arrays are 150 by 150 $\mu m$. To maintain the filling fraction, the array periods vary from 250 to 500 nm. Let us begin by considering the colour exhibited by the larger arrays (200 by 200 $\mu m$). As the nano-disk radius increases, the arrays vary in colour from blue to green, yellow and orange. Our physical interpretation of the observed colour is as follows. As discussed in relation to Figure 3(c), a band of high reflectance occurs for the spectral region between the EQ and MD resonances. As the disk radius is increased, the resonances (and therefore the high reflectance band) shift from shorter to longer wavelengths. This results in the colour change that is observed in Figure 5(d). We now consider the two smaller arrays that contain the smallest radius disks (bottom two arrays of Figure 5(d)). For these arrays, the observed colour is a combination of this band of high reflectance (between the EQ and MD) and the high reflectance that occurs on the longer-wavelength side of the MD resonance.

Dark-field optical microscopy has been previously used to measure scattering spectra from arrays of silicon nanoparticles on glass substrates,[24] where the scatterings are coupled out from diffraction ordered. Our silicon nanodisks are not isolated, however, but are fabricated in square arrays with periods ($p$) of 500 nm and below. We expect that for only wavelengths shorter than $\lambda = 2p \times NA$ could a non-zero diffraction order be excited from an array and collected by the lens. Here $NA$ denotes the numerical aperture of the lens. In other words, for longer wavelengths, the reflected field can be thought of as specular and free from higher diffraction orders. A dark-field microscope image of the arrays is shown as Figure 5(e). This image is obtained with a microscope lens with $NA = 0.25$. For wavelengths longer than 250 nm, therefore, only specular reflection is expected from the nanowire arrays. One would expect this specularly-reflected light to exhibit colour. This specularly-reflected light should not be collected by our microscope as it is a dark-field design. It can be seen however that the nanowire arrays appear brighter than their surroundings in the dark-field images of Figure 5(e). We attribute this to scattering from surface roughness and other fabrication imperfections. It is also notable that the colour variation from one array to the next is much less pronounced than for the bright-field images (Figure 5(d)), which is consistent with our expectations.

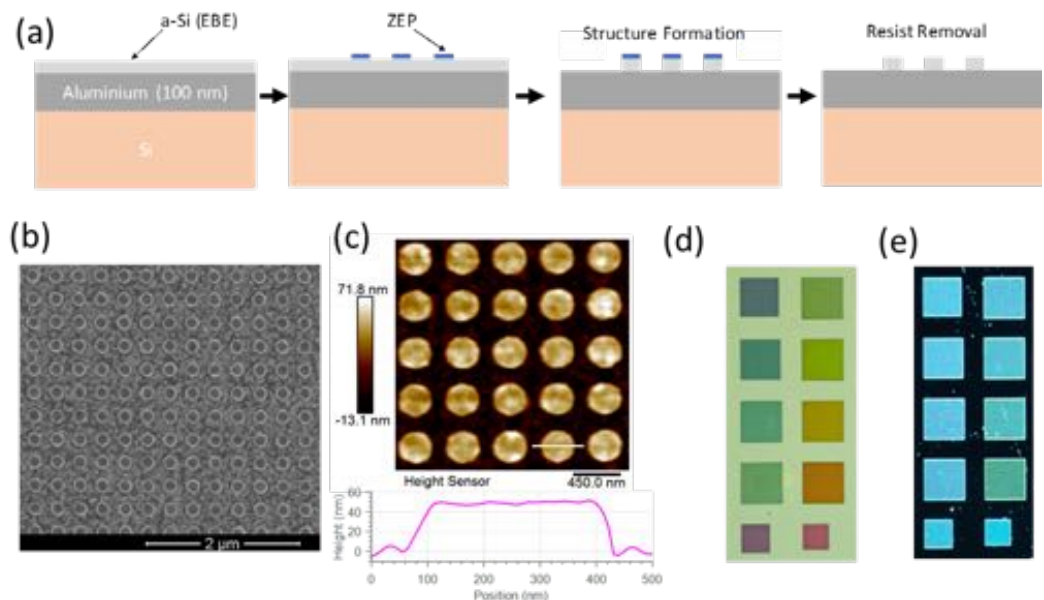

*Figure 5. Experimental demonstration of colour printing using silicon nano-disks on an aluminium substrate. (a) Schematic diagram of fabrication process. (b). Top view scanning electron microscope (SEM) image of array (radius = 90 nm; period = 300 nm). (c) Atomic force microscope (AFM) image of Si nanodisk array (radius = 135 nm; period = 450 nm), from which disk height is determined. (d). Bright-field optical microscope image of fabricated arrays. Array periods are 275, 375, 350, 325, 300, 400, 425, 450, 500, and 250 nm, clockwise from bottom left. In same sequence, the radii are 83, 113, 105, 98, 90, 120, 128, 135, 150, and 75 nm. (e). Dark field optical microscope image of sample also shown as panel d.*

We next fabricate additional samples, this time with a silicon thickness of 45 nm. In Figure 6(a), we present the reflection spectra measured from ten arrays of silicon nano-disks. The nano-disks cover the same range of radii (75 to 150 nm) and periods (250 to 500 nm) as the arrays with heights of 50 nm (Figure 5). The measured reflection spectra are shown as Figure 6(a), with the curves plotted with vertical offsets that increase with nanodisk radius. We also show bright-microscope images of the arrays that are obtained with the same microscope (and lighting conditions) that is used for Figure 5(d). These images are plotted on the right hand side of Figure 6(a). It can be seen that the colour of the background region (45 nm thick unpatterned Si film on Al), is now light blue, i.e. differing from the green colour of the 50 nm thick Si film on aluminium (Figure 5(d)). The resonance positions of the arrays are also blue-shifted, now that the disk height is decreased. For example, the array comprising 90 nm radius disks on a 300 nm period is now purplish, unlike the blue colour of the corresponding array of Figure 5(d). In Figure 6(b), we present the reflection spectra simulated for the arrays. These are in reasonable agreement with the experiments, showing a pair of dips that shift to longer wavelengths as the nano-disk radius is increased. For the array containing the nano-disks with the largest radii (150 nm), for example, the measured spectrum shows dips at ~510 nm and ~705 nm. The corresponding simulated spectrum has dips at ~500 nm and ~710 nm. As before (Figure 3(c) and (d)), there are differences between simulation and experiment however. The key differences are that the simulated spectra show an additional dip between the pair of dips discussed above, and dips are deeper (i.e. lower reflectance at the dip centre). As before, we attribute these differences to the illumination not being at purely normal incidence in experiments, and to non-idealities (e.g. surface roughness of silicon and aluminium) in fabricated samples.

As a further demonstration of colour printing, we realize the logo of our institution using small arrays of nano-disks with different radii and periods (Figure 6(c)). The outline of the shield behind the goddess figure, the name of the institution and the outline of the ribbon (containing the institutional motto in Latin) exhibit a blue colour that is produced using arrays of nano-disks with radii of $r = 90\ nm$ on a period of $p = 300\ nm$. The pink colour of the clothes and wings of the goddess is achieved using arrays of nano-disks with $r = 83\ nm$ and $p = 275\ nm$. The leaves held by the goddess have a green colour, achieved using a nano-disk array with $r = 105\ nm$ and $p = 350\ nm$. To give it a yellow tone, the skin is coloured with a nano-disk array with $r = 135\ nm$ and $p = 450\ nm$. The stars underneath the leaves are orange, achieved with nano-disk arrays with $r = 150\ nm$ and $= 500\ nm$. The letters 'POSTERA CRESCAM LAUDE' and the hair of the goddess have a dark appearance, achieved by combining disks with different radii (75 nm, 90 nm, and 105 nm).

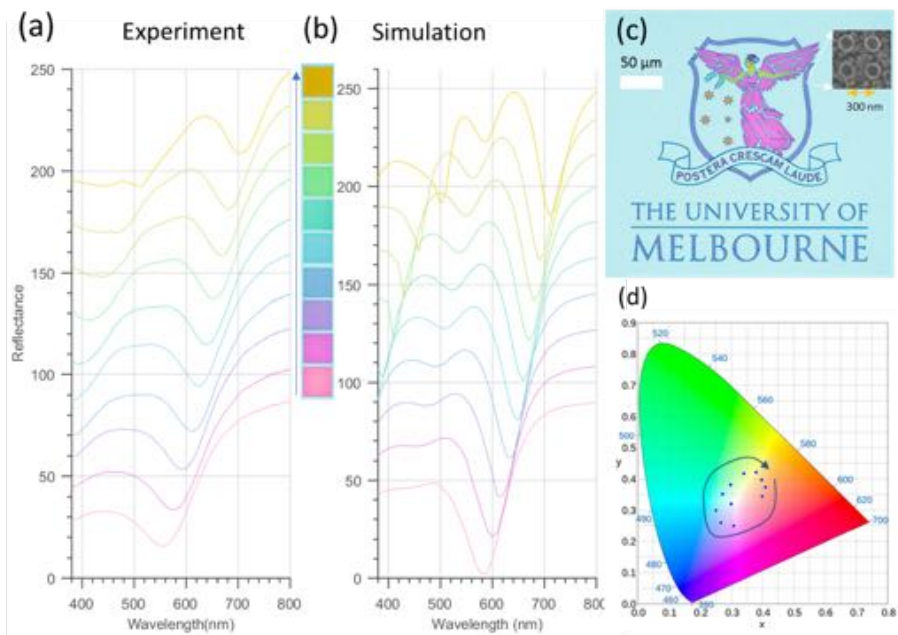

*Figure 6. Optical characterisation of colour printing using silicon nanodisks on aluminium substrate. (a) Reflectance spectra measured from arrays of silicon disks on an aluminium substrate. Inset is bright-field optical microscope image of arrays. Line colour used for plotting spectrum of each array is chosen to match colour that the array appears in optical microscope image. Successive spectra are shifted vertically by 15% per spectrum for clarity of presentation. (b). Reflectance spectra simulated for arrays with same parameters as those of panel a, using same colour coding and vertical offset schemes. (c)*

*Optical microscope image of demonstration of colour printing of institutional logo of authors of this paper. Surrounding region (light blue colour) consists of unpatterned silicon film on aluminium. (d). Positions (black dots) of measured reflectance spectra plotted in CIE 1931 chromaticity chart. Arrays increase in period as indicated by direction of arrow, in same manner as arrowed line of panel a.*

The spectra generated by the ten arrays of Figure 6(a) are converted to colour and plotted as ten colour coordinates in a standard CIE 1931 chromaticity chart as Figure 6(d). The illuminant used in this conversion is the standard D65 light source. The trajectory made by the colour coordinates of the arrays with increasing nano-disk radius is a clockwise path, from red to blue to green to orange.

The results presented thus far are for periodic arrays of nano-disks. This leads to the question of the influence of the nano-disk arrangement upon the reflection spectrum (and therefore the colour). To address this, we fabricate nano-disks in randomized arrangements at different densities. The reflection spectra measured from these devices are presented in the SI. We find that for devices with nano-disks of the same radius, the positions of the spectral dips vary with nano-disk density (as would be inevitable due to near-field coupling), but not very substantially. Earlier in this paper, we interpreted the shorter-wavelength dip as being associated with the EQ mode. As for some arrays this occurs at a wavelength similar to the array period, one wonders about the possible contribution of the lattice resonance. The results we obtain from the randomized devices confirm the importance of the EQ mode, as for these devices the shorter-wavelength dip occurs despite the absence of the lattice resonance. Optical microscope images of the devices are also presented in the SI. We find that for devices containing nano-disks with the same radius, the lower density devices have less saturated colour, which could be useful in colour printing.

**Discussion**

In this work, we propose a generalization of the method of images to allow it to be applied to the multipole expansion for dielectric nanoparticles on conductive surfaces. We anticipate that our method will enable useful physical insight into the scattering of light in this configuration, showing how the process can be described in a compact fashion by what we term the effective multipole coefficients. Our method furthermore reveals the selection rules that govern the excitation of the multipole resonances, thereby explaining which excitations are naturally forbidden by symmetry. We show how our method can be applied to real metals with loss via the concept of image strength. Lastly, we apply our method to experimentally demonstrate a new mechanism for colour printing, in which silicon nano-disks are formed on an aluminium substrate. We show that we can shift a spectral band of high reflectance across the visible spectrum by appropriate choice of nano-disk radius, thereby achieving a range of colours. This high reflectance band occurs between the spectral positions of the electric quadrupole (EQ) and magnetic dipole (MD) mode resonances. As revealed by our generalized method of images, there are other modes such as the electric dipole that would occur in this spectral range that are not excited due to symmetry considerations.

**Materials and Methods**

The effective multipole expansion was fully implemented in COMSOL Multiphysics software package and the fabrication process was described in details in Figure 5 and the associated text.

**Supplementary Materials**

Fig. S1. Concept and numerical results of dipolar mode of a particle scattering on PEC.
Fig. S2. Comparison between effectively multipole expansions and traditional multipole expansion of light scattering by an arbitrary particle on PEC by tilted incident light.
Fig. S3. Numerical results of effective multipole expansions on different conductive substrates.
Fig. S4. AFM images of periodic and randomized silicon particle arrays.
Fig. S5. Spectra and optical images of arrays with different densities.
Fig. S6. Schematics of multipole mode selection and transformation.

**Acknowledgments**

This work was supported by the Australian Research Council (DP150103736 and FT140100577) and by the Victorian Endowment for Science, Knowledge and Innovation (VESKI). Fabrication work was performed at the Melbourne Centre for Nanofabrication (MCN) in the Victorian Node of the Australian National Fabrication Facility (ANFF).

S.Q.L and K.B.C. developed the concept. S.Q.L devised the theory, performed fabrication, experiments, simulation and data analysis. S.Q.L., W.S., M. Y, and K.B.C. discussed the results. S.Q.L. and K.B.C. wrote the manuscript.

# Supplementary Information for
# "Generalized method of images and reflective color generation from ultra-thin multipole resonators"

**Authors**

S. Q. Li,[1]*† W. Song,[2,3] M. Ye[1], K. B. Crozier[1,2]*

**Affiliations**

1. Department of Electrical and Electronic Engineering, University of Melbourne, VIC 3010 Australia

2. School of Physics, University of Melbourne, VIC 3010 Australia

3. Department of Materials Science and Engineering, Huazhong University of Science and Technology, Wuhan 430030, PR China

*S. Q. Li: lishiqia@gmail.com; K. B. Crozier: Kenneth.crozier@unimelb.edu.au

† Present Address: Data Storage Institute, 2 Fusionopolis Way, #08-01 Innovis Tower, Singapore 138634

**Overview**

In the first section of this Supplementary Information document, we derive the expressions for the effective multipole coefficients that are presented in the main manuscript. These constitute the generalized method of images, an intuitive way for understanding the scattering of light by an individual nanoparticle or by an ensemble of nanoparticles on a conductive substrate. We demonstrate the application of this approach to understanding several physical phenomena that occur for scattering from nanoparticles on conductive substrates, including substrate-induced magnetic resonances. We evaluate the applicability of this method to real metals with finite conductivity, i.e. which are not perfect electrical conductors (PECs). These metals consist of gold, silver and aluminium.

In the second section, we consider the case for which the nanoparticles are not formed in regular arrays, but rather with random positions in ensembles. We study through experiments what effect this randomness has upon the colour the ensembles exhibit. We conclude that colour can be understood as arising chiefly from the properties of single disks, rather than being an outcome of the disks being formed in ordered arrays.

**Derivation of the effective multipole coefficients**

In solving scattering problems with the multipole expansion, the scattered waves are treated as comprising radiation from currents induced in the scatterer. When the object is small, this radiation can be approximated by dipole radiation.[1-4] As the object's size

2parameter increases, however, higher-order radiation sources need to be considered. This is done by decomposing the electric field scattered from the object in a basis comprising vectorial spherical harmonics about a chosen origin. This is often called the Mie or multipole expansion.[5] For arbitrarily-shaped particles, the electric field in the object induced by the incident light is treated as a volume current source that in turn generates the scattered radiation, meaning that the expansion can be performed through numerical integration.[6-8] However, the application of multipole expansions to analyze scattering from an object on a surface is still limited, as the interaction between the object and the substrate is complex, for example sometimes leading to an effect that has been termed magnetoelectric coupling. [9,10] Hence, it is of interest to the nano-optics community to explore an accurate way to describe multipole scattering from an object at an interface.[9,11-16]

In this section, starting from the Dirichlet boundary condition, we derive an orthogonal form (i.e. free of coupling among the poles) of the multipole description of scattering from an object on a perfect electrical conductor (PEC) substrate. We then generalize our method to practical substrates, i.e. those with finite conductivity. We test our formulation with gold, silver, and aluminium.

*Effective multipole coefficient for scattering by nanoparticle on a perfect electrical conductor.* We first consider the case of a silicon nanorod sitting on a PEC. This case is relatively simple, facilitating physical intuition, but is also of practical importance as many metals are reasonably approximated as PECs, especially at long wavelengths.[17] We choose to consider silicon nanorods because the high refractive index of this material at visible and near-infrared wavelengths means that, in addition to the fundamental electric dipole mode, higher order multipole resonances such as the magnetic dipole occur.[18]

The electric dipole and the magnetic dipole are the two modes of interest here.[19] They occur at similar wavelengths because they are the two lowest order modes of the multipole expansion. It can be difficult to separate these modes.[18] However, this can be achieved when the resonant particle is sitting on a PEC substrate, as we consider below. Normally-incident plane-wave illumination of a particle on a substrate induces dipoles oriented in the plane of the substrate. In the method of images, the fields generated by a dipole above a PEC are written as the sum of the fields generated by the original dipole and those generated by a virtual (image) dipole source.[20] The image dipole is located at the same distance below the PEC surface as the original dipole is above it. With this in mind, we consider the electromagnetic simulations of Figure S-1(a). The instantaneous electric field distributions are calculated for a silicon sphere in air (radius 100 nm, Figure S-1(a-I)), a silicon rod in air (radius 70 nm, height 190 nm, Figure S-1(a-II)) and a silicon rod above a PEC substrate (radius 70 nm, height 95 nm, Figure S-1(a-III)). Inspection of the field patterns of Figure S-1(a-I, II) indicates that they are very similar to those of magnetic dipole resonances, as illustrated schematically in the upper half of the Figure. We now consider the rod above the PEC. Inspection of Figure S-1(a-III) reveals that the field distribution resembles that of the upper half of the rod in air (Figure S-1 (a-II)). This is consistent with the formation of image electric and magnetic dipoles that are antiparallel and parallel to the original electric and magnetic dipoles, respectively. This leads to the cancellation of the electric dipole and the formation of a pure magnetic dipole.

The method of image dipoles for the PEC originates from the uniqueness theorem and the solution to the scalar potential for Dirichlet boundary conditions:

$$\Phi(\vec{r}) = \frac{1}{4\pi\varepsilon_o\varepsilon_r} \int_V \rho(\vec{r}')G_D(\vec{r},\vec{r}')d\vec{r}' - \frac{1}{4\pi} \int_S \Phi(\vec{r}')\frac{\partial G_D(\vec{r},\vec{r}')}{\partial n'} ds', \qquad (1)$$

The first term on the right-hand side is the potential due to the charges above the boundary, where $\rho(\vec{r}')$ is the charge density at a position $\vec{r}'$ above the PEC and $G_D(\vec{r},\vec{r}')$ is a Green's function. The second term is an integration of surface potential at the boundary. This means that the contribution to the potential from the other side of the boundary is completely specified by the potential at the boundary, regardless of the details of the charge distribution on the other side. This confirms that the method of images is rigorous. The problem then becomes that of finding the image charge distribution on the other side of the boundary that will produce the same $\Phi(\vec{r}')$ on the surface $s$ of the original problem.

We note that $\Phi(\vec{r}')$ is a constant on the PEC surface. We are then able to perform the multipole expansion to find the strength of each pole and, using the boundary condition, the strengths of the corresponding image poles. We choose the origin of the expansion to be on the PEC surface, so the source and image poles will be co-located. The overlapped origins facilitate a definition of what we call the **effective multipole coefficients**, which is a set of coefficients for point multipole sources with the image multipoles taken into account.

We start by analyzing the magnetic field generated by individual magnetic multipole sources. For an arbitrary multipole source, the electric field due to magnetic multipole is expressed as $\vec{E}^m(l,m) = Zg(l)\vec{L}Y(l,m)$, where $Z$ is the impedance of the medium, $g(l)$ is the radial component of the solution of order $l$, $Y(l,m)$ is the spherical harmonic of order $l$ and degree $m$, and $\vec{L}$ is the orbital angular-momentum operator, defined as $\vec{L} = \frac{1}{i}(\vec{r}\times\nabla)$. We can expand the term on the right-hand side of the equation to write the electric field in Cartesian coordinates explicitly,

$$\vec{E}^m(l,m) = Zg(l)\begin{pmatrix} 0.5\left(\sqrt{(l+m)(l-m+1)}Y(l,m-1) + \sqrt{(l-m)(l+m+1)}Y(l,m+1)\right)\hat{x} \\ 0.5i\left(\sqrt{(l+m)(l-m+1)}Y(l,m-1) - \sqrt{(l-m)(l+m+1)}Y(l,m+1)\right)\hat{y} \\ mY(l,m)\hat{z} \end{pmatrix}$$

, where the spherical harmonic terms ($Y(l,m-1)$ and $Y(l,m+1)$) are the only terms with angular dependence (i.e. have θ and φ as independent variables). The Dirichlet boundary condition requires that the *y* and *z* components of the electric field vanish at φ = π/2 and 3π/2, for the choice of coordinates in which the PEC lies in the *y* − *z* plane and the *x* axis is normal to its surface. With this condition, we find that the image pair term of the magnetic multipole term (*l*, *m*) is the term (*l*, -*m*).

We now move forward to find the image sources for the electric multipoles. Instead of finding the electric field directly for electric multipoles, it is simpler to evaluate the scalar product between the unit position vector $\vec{r}$ and the electric field $\vec{E}$ generated by each multipole source. The Dirichlet condition implies that $\vec{r} \cdot \vec{E}$ needs to be zero when $\vec{r}$ is parallel to the PEC boundary, i.e. when $\varphi = \pi/2$ or $3\pi/2$ (Figure S-2). Furthermore, the electric multipole field due to each pole satisfies the following equation

$$\vec{r} \cdot \vec{E}^e(l,m) = -Z\frac{l(l+1)}{k}f(l)Y(l,m),$$

where the radial dependence is included in the term $f(l)$. Similar to magnetic multipoles, we find that the image pair term of the multipole with the order *(l, m)* is the term *(l,-m)*, but of opposite sign.

We name a new set of multipole coefficients, termed the **effective multipole coefficients**, by taking the image multipoles into consideration. We express these in terms of the original multipole coefficients using the following formulae,

$$a^e_{eff}(l,m) = -a^e_{eff}(l,-m) = a^e(l,m) - a^e(l,-m) \tag{2}$$

$$a^m_{eff}(l,m) = a^m_{eff}(l,-m) = a^m(l,m) + a^m(l,-m) \tag{3}$$

As an example, the Cartesian dipole moments can then be expressed with the multipole coefficients obtained as follows,[21]

$$p_x = C_1\left(a^e_{eff}(1,-1) - a^e_{eff}(1,1)\right) = 2C_1\left[a^e(1,-1) - a^e(1,1)\right]$$
$$p_y = -iC_1\left(a^e_{eff}(1,-1) + a^e_{eff}(1,1)\right) = 0$$
$$p_z = \sqrt{2}a^e_{eff}(1,0) = 0$$
$$m_x = cC_1\left(a^m_{eff}(1,-1) - a^m_{eff}(1,1)\right) = 0$$
$$m_y = cC_1\left(a^m_{eff}(1,-1) + a^m_{eff}(1,1)\right) = -icC_1\left[a^m(1,1) + a^m(1,-1)\right]$$
$$m_z = \sqrt{2}cC_1 a^m_{eff}(1,0) = 2\sqrt{2}cC_1 a^m(1,0)$$

, where $C_1 = \dfrac{-3\pi E_i}{ik^3}$ and $E_i$ is the incident field strength. These results agree with the method of images for dipoles. [20,21]

In Figure S-1(b), we plot the contributions to the scattering cross-section from ED and MD components for the nanorod in air (full length, i.e. height is 190 nm) and for the nanorod sitting on the PEC surface (half-length, i.e. height is 95 nm). The contributions of different poles are calculated as follows:[6]

$$C^{e,m}(l) = \frac{\pi}{k^2}\sum_{m=-l}^{l}(2l+1)|a^{e,m}(l,m)|^2$$

, where effective multipole coefficients are used in the case of the nanorod on the PEC. The total cross-section is the incoherent summation of all orders of multipoles, as the basis of the effective multipole expansion is the same as the orthogonal basis of the direct multipole expansion.

For comparison, we calculate the scattering cross-section ($C_{Scat}$) by integrating the Poynting vector of the scattered field at the surface of the nanorod,

$$C_{Scat} = \frac{\int \vec{S}_s \cdot \hat{n} da}{I} = \frac{\int \text{real}(\vec{E}_s \times \vec{H}_s^*) \cdot \hat{n} da}{\frac{1}{2}\sqrt{\frac{\varepsilon_o \varepsilon_r}{\mu_o}}|E_i|^2}$$

, where $\vec{S}_s$ is the time-averaged Poynting vector of the scattered fields on the surface of the particle, $\hat{n}$ is the surface normal, and $da$ is the differential surface area. The scattered electric field $\vec{E}_s$ is defined as the difference between total electric field $\vec{E}_t$ and the incident electric field $\vec{E}_i$ (i.e. $\vec{E}_s = \vec{E}_t - \vec{E}_i$). The scattered magnetic field $\vec{H}_s$ is defined in the same way (i.e. $\vec{H}_s = \vec{H}_t - \vec{H}_i$). The incident electric and magnetic fields are those that would occur in the absence of the particle. $I$ is the intensity of the incident light. It can be seen from Figure S-1(b) that there is good agreement between the scattering cross section found by direct integration, and that found by considering the contributions of the ED and MD. This confirms the accuracy of our approach. The multipole coefficients are obtained by the integration of the scattering current in the particle, which is defined as $\vec{J}_s = -i\omega\varepsilon_o(\varepsilon_r - \varepsilon_1)\vec{E}_t$, where $\varepsilon_r$ and $\varepsilon_1$ are the relative permittivities of the particle and of the medium in which it is embedded, respectively.[6-8]

To test the generality of our method, we reduce the symmetry by considering the case where the incident plane wave is tilted to an angle of 60 degrees from the normal. We also rotate the cylindrical nanorod about a randomly chosen axis (Figure S-2(a)). The sum of the effective multipole coefficients (dipoles and quadrupoles only) still matches perfectly with the scattering cross section $C_{Scat}$ found by direction integration of the scattered field (Figure S-2(b)). On the other hand, the total scattering cross-section calculated from the original (traditional) multipole expansion deviates significantly from $C_{Scat}$, shown in Figure S-2(c). This deviation can be explained as the quenching of radiation sources by the PEC boundary, which is not taken into account by the original multipole expansion.

*Extension to conductive surfaces with the loss.* We next consider the case of the substrate having finite conductivity (i.e. not a PEC). We must therefore modify the virtual radiation sources to take into account losses in the substrate. We postulate the following generalized formulae,

$$a_{eff}^e(l,m) = a^e(l,m) - \frac{\varepsilon_s - \varepsilon_1}{\varepsilon_s + \varepsilon_1} a^e(l,-m) \tag{4}$$

$$a_{eff}^m(l,m) = a^m(l,m) + \frac{\varepsilon_s - \varepsilon_1}{\varepsilon_s + \varepsilon_1} a^m(l,-m) \tag{5}$$

where $\varepsilon_s$ is the relative permittivity of the substrate and $\varepsilon_1$ is the relative permittivity of the medium in which the scattering particle is situated. It can be seen that these equations are an extension of the image dipole approximation for a planar interface (see e.g. Ref [22]). These equations are consistent with Equations (2) and (3). That is, as $\varepsilon_s$ approaches $-\infty$ in the limit in Equations (4) and (5), i.e. for a PEC substrate, we recover Equations (2) and (3). Furthermore, the original result for multipole expansion in a homogeneous environment is obtained when we set $\varepsilon_s = \varepsilon_1$, as the second terms on the right-hand sides of Equations (4) and (5) are then equal to zero. We conclude that this approach behaves well asymptotically. Note that in the latter case (i.e. setting $\varepsilon_s = \varepsilon_1$), the origin of the multipole expansion is at the edge of the scattering object, which is not favorable as it is not at the center of symmetry. We denote $\frac{\varepsilon_s - \varepsilon_1}{\varepsilon_s + \varepsilon_1}$ as $\Gamma$, termed the image strength.

Using the above prescription, we calculate light scattering from a silicon sphere. The permittivity values of gold, silver, and aluminium used here are from Palik.[23] In Figure S-3(a-c), we show the contributions to the scattering cross-section from the multipole components using Equations (4) and (5) for gold, silver and aluminium, respectively. We furthermore show the sum of the scattering cross section contributions from the multipoles, up to the octopole terms. The sum of multipole scattering cross sections is in general agreement with $C_{Scat}$ computed by direct integration of the fields on the nanoparticle surface. For each metal considered, both curves show two spectral peaks, whose locations are in good agreement. It can be seen however that for all of the metals, the multipole scattering sum is smaller than $C_{Scat}$. The contributions of the octopoles are small, and we thus believe that the fact that the higher multipoles are not included in the summation is unlikely to be the reason for the difference between the two approaches. However, we note that some of the field scattered by the particle is absorbed by the metal substrate and not radiated to the far field. This is taken into account by the effective multipole method but ignored when directly integrating the Poynting vector on the surface of the nanoparticle. Comparison between the sum of multipole contributions and the scattering cross-section calculated from direct integration of the Poynting vector on a hemisphere whose radius is one wavelength ($C_{Scat, farfield}$) shows better agreement.

For comparison, we also calculate the PEC case (Figure S-3(d)). It can be seen that the results for the PEC are in broad agreement with those for the real metals (Figure S-3(a)-(c)), but with some differences that are most pronounced for the gold substrate. The resonances of the latter are significantly broader than the PEC case. The electric quadrupole mode is significant and peaks at $\lambda = 470\ nm$ for the PEC case, as indicated by the dashed vertical line in Figure S-3(d). For the gold substrate case, the electric quadrupole mode is weaker and is red-shifted (again indicated by dashed vertical line, Figure S-3(a)). The instantaneous electric field distribution for the PEC-substrate case is shown as Figure S-3(e). This is calculated for illumination at the wavelength of the electric quadrupole resonance peak, as denoted by both the vertical dashed line and the letter "A" in Figure S-3(d). It can be seen that the field distribution has a quadrupole nature, and is complemented by its image pole.

For the gold substrate case, there is some electric dipole component in the spectral range between 450 nm and 500 nm that is not present in the PEC case. We note that these differences between the gold and PEC cases are at spectral locations for which the gold has high loss and where the image strength has a large imaginary component (Figure S-3(f)). It is further confirmed by that the broadening effect is less in silver and in aluminium than in gold. In the case of aluminium, the multipole expansion result agrees well with $C_{Scat}$. This is consistent with the fact that, for aluminium, the image strength $\Gamma$ is approximately unity across the visible spectrum (Figure S-3(f)).

In summary, we formulated a multipole expansion method to account for the scattering of a particle or an ensemble of particles on a conductive substrate. We verified the physical validity of this method by considering individual particles and a particle array. We furthermore extended our approach beyond perfect electrical conductor substrates to those with finite conductivity. In particular, we examined the predictions of our model for gold, silver and aluminium substrates at visible wavelengths. The effective multipole coefficients we obtained were consistent with the scattering cross sections predicted by integration of the Poynting vector. We found that the best agreement is obtained for the case an aluminium substrate. Our multipole expansion method is based on considering the fields internal to the particle as being a scattering current source. Provided that these internal fields are known, further consideration of the illumination conditions is unnecessary. Our method can thus be readily applied to nanoparticles in complex environments, including in random or ordered ensembles of other nanoparticles.[24,25]

**Design, Fabrication, and Evaluation of Randomized Arrays**
The nano-disks we present in the main manuscript are in square arrays whose periods are 500 nm and smaller. Such arrays are relatively straightforward to fabricate and to simulate. However, this leads to the question of whether lattice modes are important in determining the colour that an ensemble of nano-disks exhibit. To address this question, we here present studies of randomized arrays and compare them with their periodic counterparts. We conclude that the contribution of lattice modes to the colour of the arrays is not substantial.

The fabrication process for the randomized arrays is the same as that used for the ordered arrays, as presented in the main manuscript. The pattern used in the electron beam lithography step is produced as follows. We begin by generating square arrays of nano-disks with five different filling fractions, ranging from 0.05 to 0.25 in steps of 0.05. Each disk in the each array is then shifted along the two principal axes of the array by a random step size that is uniformly distributed over the range $-p/2$ to $p/2$, where $p$ is the pitch of the original square array.

Atomic force microscope (AFM) images of ordered (square lattice) and randomized nano-disk arrays are shown as Figure S-4(a) and S-4(b), respectively. The Fourier tranforms (FTs) of these AFM images are shown as Figure S4(c) and S4(d). It can be seen that the FT pattern of the ordered array exhibits diffraction spots, while the pattern from the randomized array only possesses broad ring patterns. This implies that these arrays have short range order, which is consistent with the algorithm used to generate them.

In Figure S4(e) to S4(g), we show AFM images of arrays of three different densities and their FT patterns. Due to the fact that, for the high density arrays, the disks are shifted by smaller distances, the array whose filling fraction is 0.25 is more ordered. Nevertheless, the FT pattern shows no diffraction spots.

In Figure S-5, we show the reflectance spectra measured for the randomized arrays of disks with different filling fractions, together with images obtained with a bright-field optical microscope. We also plot the spectra and images for the ordered arrays under the same lighting conditions and microscope settings. Five different sets of arrays are shown. The nano-disks vary in radius from 75 nm to 150 nm, in increments of 15 nm. It can be seen that the positions of the reflectance dips are consistent between arrays of nano-disks of the same radius at different filling fractions. It can be seen from the optical microscope images that the colour saturation of the arrays decreases as the filling fraction decreases. It can also be seen that the resonances red-shift as the filling fraction decreases, a consequence of the weaker near-field coupling between nano-disks. From the spectra and visual appearance of the arrays under the microscope, we can conclude that the lattice resonances as shown in the main text do not play a significant role in determining the colour. Because the colour is not a lattice resonance effect, the colour saturation of the arrays can be varied by changes to the filling fraction.

**Understanding the selection and re-ordering of multipoles that occurs with the introduction of the PEC using the generalized method of images**

As discussed in the main manuscript, when a scatterer is placed on a PEC surface, the multipoles can experience different effects. Some multipoles can be eliminated, while others are unaffected. In addition, a resonance that occurs in the response of a certain multipole in the isolated scatterer case can appear in the response of a different multipole when the scatterer in placed on a PEC surface. Here we provide some physical insight into this process by considering a few examples of multipolar excitations (a quadrupole and an octupole) and breaking them down into electric current sources.[26] We begin by considering a quadrupole comprising two antiparallel time-harmonic electric current sources (red arrows of Figure 6-S(a)). We assume that these currents are located infinitesimally close to the surface of the PEC. This is consistent with our method, as it integrates the current sources about an origin that is on the PEC. It can be seen that the combination of the currents and their images will not radiate, as each current source cancels with its image current. This implies that these quadrupole mode will be eliminated with the presence of the PEC. We next consider an octupole (Figure 6-S(b)). It can be seen that some of the current sources orient in the plane of PEC but some of the sources orient out of the plane. The in-plane current sources cancel with their images, while the out-of-plane sources interfere constructively with their images. The resulting radiation pattern is then as if it is from a quadrupole polarized out of the plane.

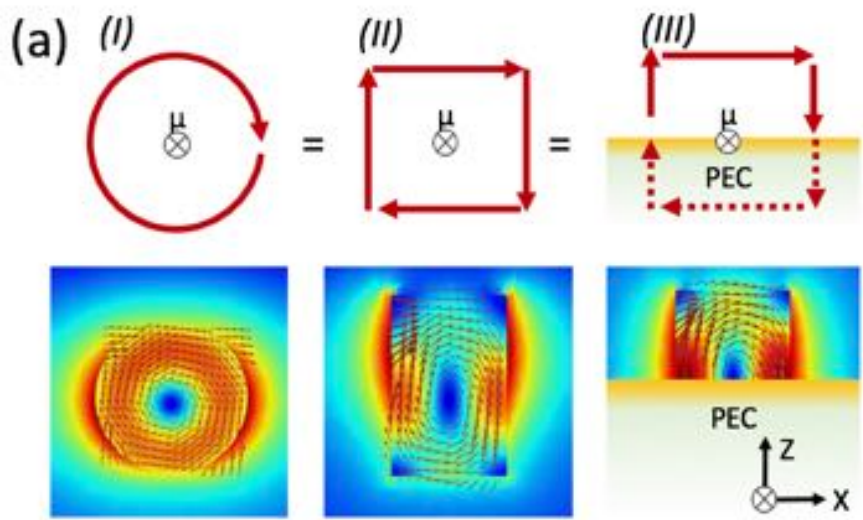

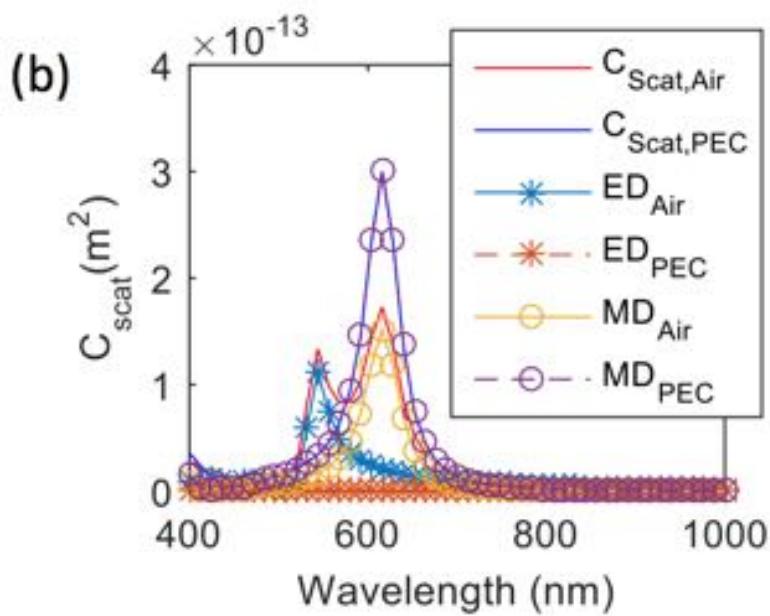

Figure S-1. **Concept and numerical results of dipolar mode of a particle scattering on PEC.** (a) Schematic illustrations (upper panels) and field distributions (lower panels) of magnetic resonances induced by normal incidence illumination (propagation direction: -$x$) of (I) Si sphere (100 nm radius, at wavelength of $\lambda = 772\ nm$) in air, (II) Si nanorod (70 nm radius, 190 nm height, at wavelength of $\lambda = 590\ nm$) in air, (III) Si nanorod (70 nm radius, 95 nm height, at wavelength of $\lambda = 590\ nm$) on PEC. Illumination is polarized in $z$ direction. Field distributions are found by finite element modelling. Arrows and colours show direction and magnitude of instantaneous electric field, respectively. (b) Partial scattering cross sections for ED and MD components, calculated for the configurations (II) and (III) of panel a. Total scattering cross section is also plotted.

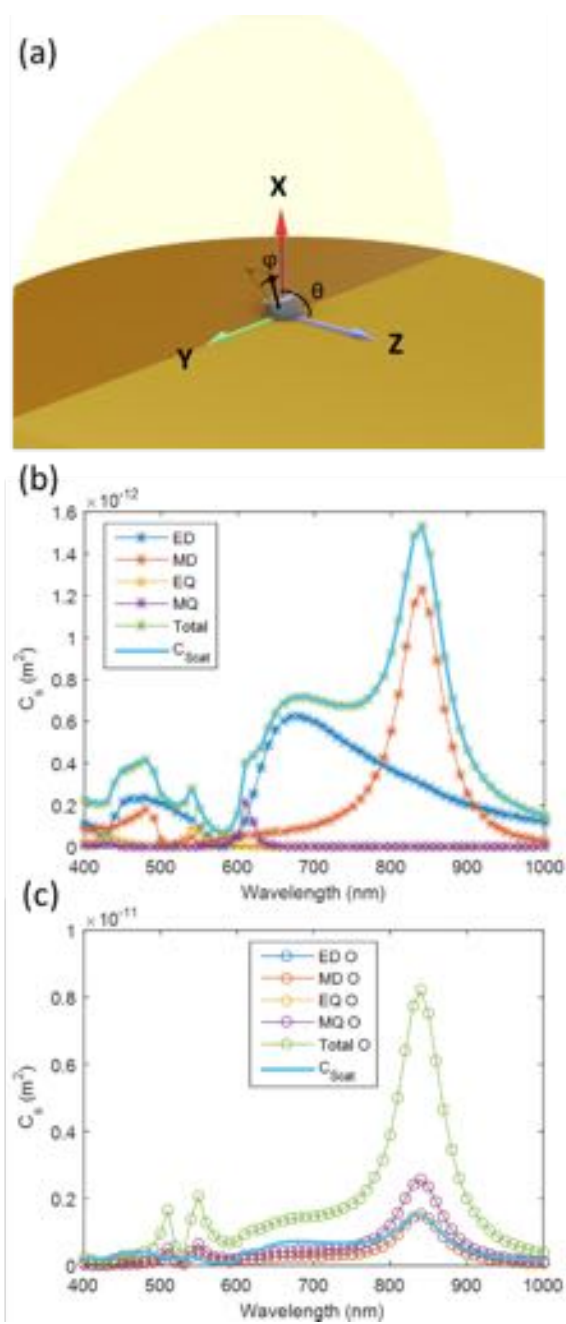

Figure S-2. **Comparison between effectively multipole expansions and traditional multipole expansion of light scattering by an arbitrary particle on PEC by tilted incident light.** (a). Schematic illustration of coordinate system chosen in this work. PEC boundary is *yz* plane, depicted as brown-yellow surface. Illumination is at 60 degrees from the normal, with plane of incidence being *xy* plane. Grey object at origin of coordinate system is part of an arbitrarily-oriented nanorod. Orientation of the nanorod is as follows. Nanorod axis is originally along x-direction, but is then nano-rod is rotated by 27 degrees about an axis that is defined by a vector that includes the origin and the point with coordinates (0.76, 0.4, 0.1). (b). Partial scattering cross sections for the effective multipole terms and their summation. Total scattering cross section found by Poynting

vector integration is also shown. (c). Partial scattering cross sections found for the multipole terms by the original (traditional) method and their summation. Total scattering cross section (via Poynting vector integration) is also shown.

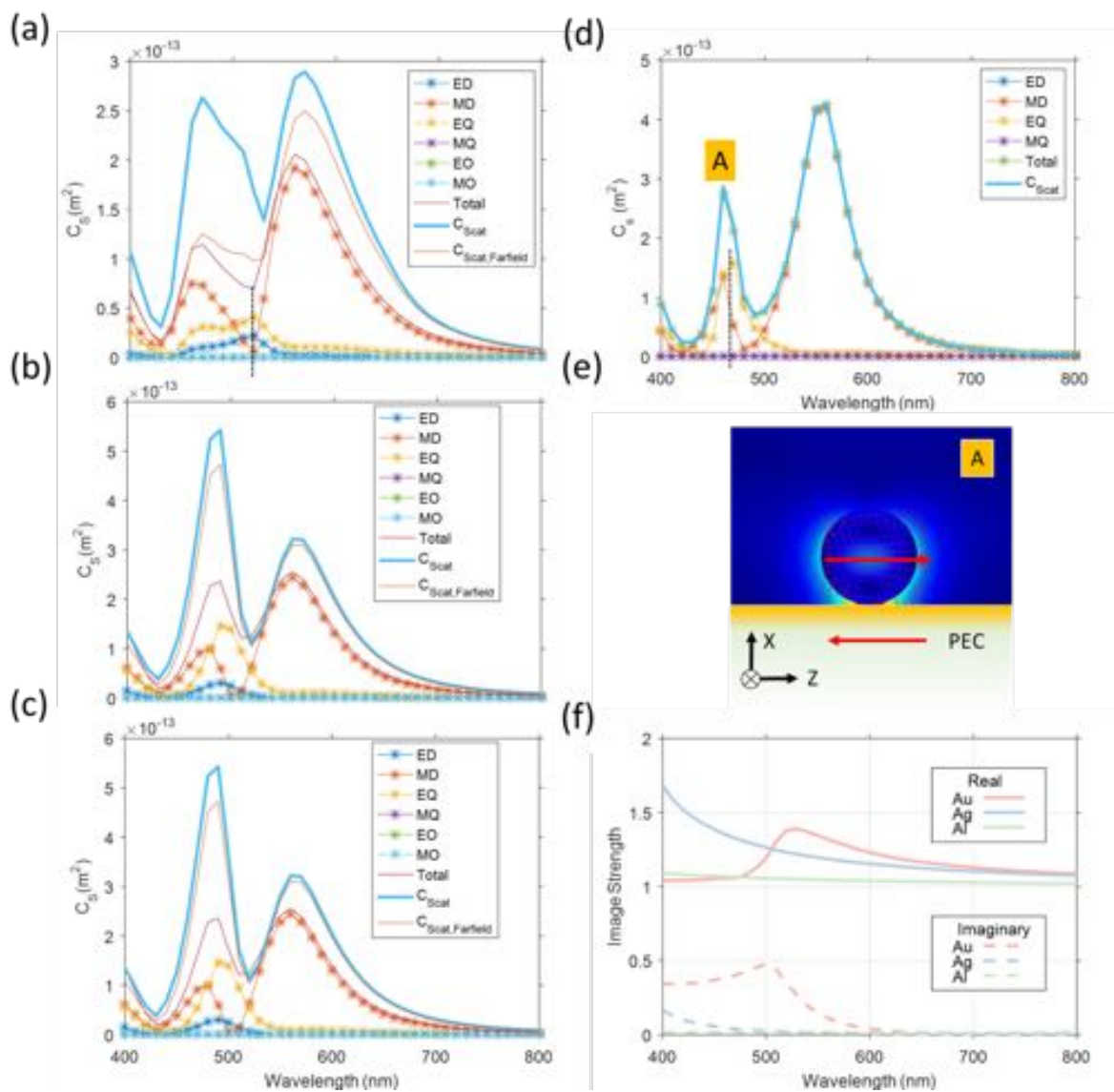

Figure S-3. **Numerical results of effective multipole expansions on different conductive substrates.** (a-d) Contributions to scattering cross-section from different multipoles (up to octopoles, $l = 3$) for silicon nanosphere (radius is 65 nm) on (a). gold, (b) silver, (c) aluminium, and (d) PEC surfaces. Red solid curve ("total"): summation of multipole contributions to scattering cross-section. Thick light blue solid curve ("$C_{scat}$"): scattering cross section found by integration of Poynting vector over nanosphere surface. Orange solid curve ("$C_{scat,\ Farfield}$"): scattering cross section found by direct integration of Poynting vector over hemisphere with radius equal to one wavelength. (e) Electric field distribution for nanosphere on PEC, calculated for normal incidence illumination at wavelength of quadrupole peak (denoted "A") of panel d. (f) Real and imaginary parts of image strength $\Gamma$ as a function of wavelength for the three different metals.

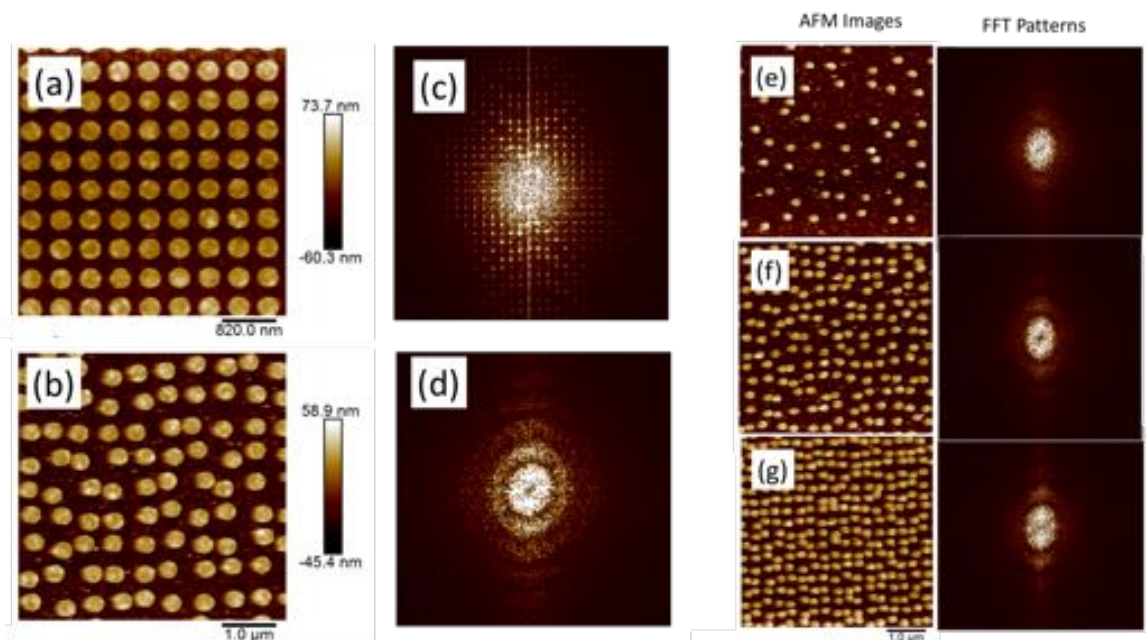

Figure S-4. **AFM images of periodic and randomized silicon particle arrays.** (a). Atomic force microscope (AFM) image of ordered array of silicon nano-disks on aluminium substrate. Nano-disks have radii of 150 nm and are on a square array of pitch 500 nm, so the filling fraction is FF~0.2827. (b). AFM image of randomized array (FF=0.2). Nano-disks have radii of 150 nm. (c). Fourier transform (FT) of array of panel a. (d). FT of array of panel b. (e)-(g). AFM images (left) and FT patterns (right) of randomized arrays with three different filling fractions (0.05, 0.15, 0.25).

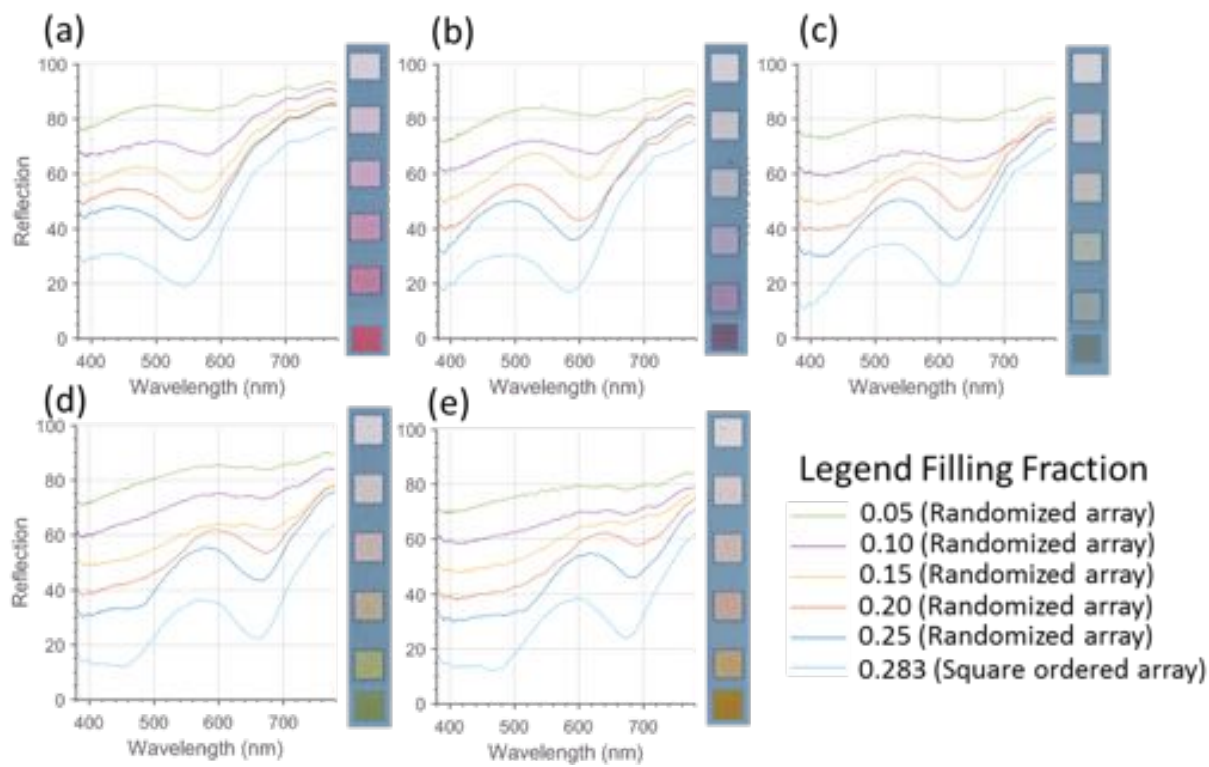

Figure S-5 **Spectra and optical images of arrays with different densities.** In each of the sub-figures from (a) to (e), we plot the reflectance spectra of arrays of randomly distributed nano-disks with five different filling fractions (FF ranges from 0.05 to 0.25, in steps of 0.05) together with the reflectance spectrum from an ordered array with F~0.283. The radii of the disks increase from (a) to (e), from 75 nm to 150 nm, in steps of 15 nm. In the insert of each sub-figure, a bright-field optical microscope image of the arrays is shown. The array FFs increases from top (0.05) to bottom (0.283).

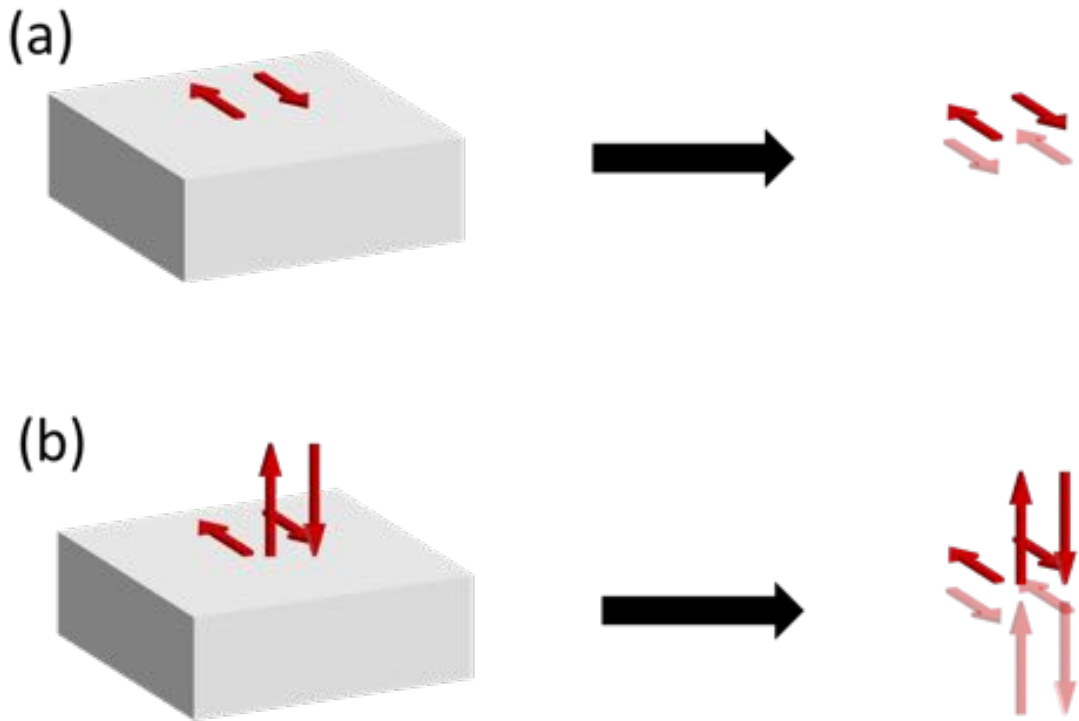

Figure S-6 **Schematics of multipole mode selection and transformation.** (a) Illustration of a quadrupole configuration that is supported in free space, but is eliminated with the inclusion of the PEC. (b) Illustration of a current configuration that is octupolar in free space, but strongly modified with the inclusion of the PEC. It can be seen that the in-plane currents are cancelled by their images, while the out-of-plane components interfere constructively with their images. This transforms the radiation pattern into being that of a quadrupole. The grayish blocks on the left panels denote the semi-infinite PEC. The semi-transparent arrows on the right panels represent the image currents.